\documentclass[aps,prd,twocolumn,nofootinbib,citeautoscript,showkeys]{revtex4-1}        
\usepackage{amsmath,amssymb,mathrsfs,bm,setspace}
\usepackage{graphicx}  
\usepackage[tight]{subfigure}     
\usepackage{color} 
\usepackage{braket}
\usepackage{array}
\usepackage[all]{xy}

\usepackage[colorlinks=true]{hyperref}  
\hypersetup{
    bookmarks=true,         
    unicode=false,          
    pdftoolbar=true,        
    pdfmenubar=true,        
    pdffitwindow=false,     
    pdfstartview={FitH},    
    pdftitle={My title},    
    pdfauthor={Author},     
    pdfsubject={Subject},   
    pdfcreator={Creator},   
    pdfproducer={Producer}, 
    pdfkeywords={keyword1} {key2} {key3}, 
    pdfnewwindow=true,      
    colorlinks=true,       
    linkcolor=blue, 
    citecolor=blue,        
    filecolor=magenta,      
    urlcolor=cyan           
} 
\definecolor{darkblue}{rgb}{0.2, 0, 0.8}
\definecolor{darkgreen}{rgb}{0.2, 0.71, 0}

\numberwithin{equation}{section}

\newcommand{\be}{\begin{equation*}}
\newcommand{\ee}{\end{equation*}}
\newcommand{\ben}{\begin{equation}}
\newcommand{\een}{\end{equation}}
\newcommand{\beqa}{\begin{eqnarray*}}
\newcommand{\eeqa}{\end{eqnarray*}}
\newcommand{\beqan}{\begin{eqnarray}}
\newcommand{\eeqan}{\end{eqnarray}}

\newcommand{\R}{\mathcal{R}}
\newcommand{\Z}{\mathcal{Z}}

\newcommand{\E}{\mathcal{E}}

\newcommand{\cM}{\mathcal{M}}

\renewcommand{\href}[2]{#2}

\def\Sol{\mathrm{Sol}}

\def\cS{\mathcal{S}}
\def\Sp{\mathrm{Sp}}
\def\cM{\mathcal{M}}
\def\bR{\mathbb{R}}
\def\PSL{\mathrm{PSL}}
\def\Iso{\mathrm{Iso}}
\def\SL{\mathrm{SL}}
\def\SU{\mathrm{SU}}
\def\E{\mathrm{E}}
\def\cl{\mathrm{cl}}
\def\Aut{\mathrm{Aut}}
\def\ad{\mathrm{ad}}
\def\cN{\mathcal{N}}
\def\cG{\mathcal{G}}
\def\fM{\mathfrak{M}}
\def\cC{\mathcal{C}}
\def\cU{\mathcal{U}}

\def\R{\mathbb{R}}
\def\Z{\mathbb{Z}}
\def\eqdef{\stackrel{\mathrm{def}}{=}}
\def\U{\mathrm{U}}
\def\mS{\mathrm{S}}
\def\SO{\mathrm{SO}}
\def\cF{\mathcal{F}}
\def\C{\mathbb{C}}
\def\H{\mathbb{H}}
\def\dd{\mathrm{d}}
\def\Im{\mathrm{Im}}

\def\Hom{\mathrm{Hom}}

\def\eff{\mathrm{eff}}

\begin{document}

\title{Geometric U-folds in four dimensions}  
\author{C. I. Lazaroiu and C. S. Shahbazi}\vspace{0.1cm}
\affiliation{Center for Geometry and Physics, Institute for Basic
  Science, Pohang, Republic of Korea 37673 } 
\email{calin@ibs.re.kr} 
\affiliation{Institut de Physique Th\'eorique, CEA-Saclay, France.}
\email{carlos.shabazi-alonso@cea.fr}

\date{\today}

\begin{abstract}     
We describe a general construction of geometric U-folds compatible
with a non-trivial extension of the global formulation of four-dimensional extended supergravity on a differentiable spin manifold. The topology of geometric U-folds depends on certain flat fiber bundles which encode how supergravity fields
are globally glued together. We show that smooth non-trivial U-folds of this type
can exist only in theories where both the scalar and space-time
manifolds have non-trivial fundamental group and in addition the
scalar map of the solution is homotopically non-trivial. Consistency with string theory requires smooth geometric U-folds to be glued using subgroups of the effective discrete U-duality group, implying that the
fundamental group of the scalar manifold of such solutions must be a
subgroup of the latter. We construct simple examples of geometric
U-folds in a generalization of the axion-dilaton model of $\cN=2$
supergravity coupled to a single vector multiplet, whose scalar
manifold is a generally non-compact Riemann surface of genus at least two
endowed with its uniformizing metric. We also discuss the relation
between geometric U-folds and a moduli space of flat
connections defined on the scalar manifold, which involves certain
character varieties not studied in the literature.
\end{abstract}   

\maketitle \singlespacing

\section{Introduction and main results} 
\label{sec:intro}  

U-folds are consistent backgrounds that incorporate in a non-trivial
manner the natural symmetries of a supergravity or string theory
\cite{Vafa:1996xn,Kumar:1996zx,Meessen:1998qm,Hellerman:2002ax,Dabholkar:2002sy,Fidanza:2003zi,Hull:2004in}. In this note, a U-fold means a {\em supergravity} background which can be obtained by gluing local solutions using U-dualities, aside from local
diffeomorphisms and gauge transformations. Such solutions can be
promoted to string theory backgrounds only when all U-dualities
involved belong to the discrete subgroup allowed by charge
quantization.

Many particular constructions of U-folds have been considered in the
literature (see, for example,
\cite{McOrist:2010jw,Cederwall:2007je,Grana:2008yw,Andriot:2012wx,
  Blumenhagen:2013aia,Braun:2013yla,Sakatani:2014hba,Candelas:2014jma,Candelas:2014kma,Cederwall:2014opa,Lust:2015yia}),
where it was often suggested that some of them do not admit any
geometric description. However, no fully general and precise
mathematical definition of U-folds has yet been given. Due to this
fact, it is unclear which U-folds may admit equivalent (though
possibly non-standard) descriptions through ordinary objects of
differential geometry, namely objects obtained via constructions
involving manifolds and smooth maps satisfying various conditions ---
the latter of which include fiber bundles. It is thus possible that
many backgrounds currently postulated to be ``non-geometric'' may in
fact admit descriptions within the framework of {\em global}
differential geometry --- albeit such a description may be
``non-standard''.

References \cite{Heterotictwist,Shahbazi:2015sba} considered a
particular construction of non-standard geometric solutions based on a
large class of non-simply-connected complex manifolds with properties
that are quite different from those of traditional supersymmetric
backgrounds. It was argued that those solutions, though constructed
geometrically in terms of manifolds and bundles, can be interpreted as
U-folds and hence {\em appear} to be ``non-geometric'' when viewed
from the perspective of patching traditional local solutions using an
open cover, in the sense that the gluing of the restrictions of the
global solution to the open sets of a cover involves non-trivial
U-duality transformations. This shows how it is possible to construct
large classes of apparently non-geometric backgrounds using ordinary
manifolds and bundles, provided that globally the geometric objects
involved are topologically non-trivial.

In this note we propose a general geometric construction of a class of
U-folds (called below {\em geometric U-folds}), which can be
considered in any four-dimensional extended supergravity theory. The
general global formulation of such theories on a space-time manifold
$M$ requires one to specify a certain flat symplectic vector bundle
$\cS$ defined over the target manifold $\cM$ of the kinetic sigma
model of scalar fields. The structure group of $\cS$ is the U-duality
group $G_0$ of the theory acting in an appropriate symplectic
representation $\rho$ which encodes electric-magnetic dualities. More
precisely, $\cS$ is the vector bundle associated through $\rho$ to a
flat principal $G$-bundle $Q$ defined over $\cM$, where $Q$ must be
specified when defining the theory. Given a classical solution $\Sol$
of the equations of motion with underlying space-time $M$, the
pull-back of $\cS$ through the sigma model map $\Phi:M\rightarrow
\cM$, which encodes the scalar fields of $\Sol$, gives a flat vector
bundle $\cS_\Phi$ defined over $M$. When restricted to the open sets
of a trivializing cover, the transition functions of $\cS_\Phi$ encode
U-duality transformations. The crucial observation is that the flat bundle
$\cS$ can be topologically non-trivial, hence $\cS_\Phi$ can also be
non-trivial provided that $M$ and $\cM$ are not simply-connected and
that the homotopy class of $\Phi$ differs from that of the constant
map. In this case, at least one of the transition functions of
$\cS_\Phi$ must be non-trivial, so $\Sol$ can be interpreted as a
non-trivial U-fold. As remarked in reference \cite{Hull:2004in}, a non-trivial space-time fundamental group is a standard underlying assumption in the construction of non-trivial U-folds. In this note, we are able to show in precise mathematical terms what is the general relation between the fundamental group of the space-time manifold and the global topology of a given U-fold solution. In fact, we obtain that such relation involves in a crucial way the fundamental group of the scalar manifold and the homotopy class of the scalar map in a very specific way. To the best knowledge of the authors, these are some of the first generic results on the global topology of U-folds available in the literature. 

Under the same assumptions on $M$, $\cM$ and
$\Phi$, the U-fold interpretation applies even for fluxless solutions
(solutions for which all electromagnetic field strengths are
identically zero), since in that case $\Sol$ can be viewed as a {\em
  multivalued} solution of the standard supergravity theory (the
theory constructed using the universal cover $\cM_0$ of $\cM$), which
is glued from local solutions of the latter using U-duality
transformations.  Thus a global geometric solution can {\em appear }to
be ``non-geometric'' when interpreted by patching its restrictions to
the open sets of a cover, the reason being, as in
\cite{Heterotictwist,Shahbazi:2015sba}, that the global solution
involves topologically non-trivial geometric objects. As familiar from
the cosmic string literature, non-simply-connected spacetimes can
often be mimicked by considering source-full solutions defined on
simply-connected spacetimes containing localized codimension two
sources. In such set-ups, the regular part of the full solution is a
source-free solution defined on the complement of all localized
sources, a complement which is incomplete and need not be simply
connected. In this paper, we consider source-free solutions (which may
be restrictions of source-full solutions to the complement of all
localized sources).

The standard formulation of four-dimensional supergravity theories
involves scalar manifolds $\cM_0$ which are simply-connected; for
$\cN\geq 3$, these manifolds are symmetric spaces of non-compact type,
which are in fact contractible. As a consequence, the flat bundles
$Q_0$ and $\cS_0$ of the standard formulation are always
trivial. However, the local computations leading to the construction
of supergravity theories only fix the scalar manifold $\cM$ up to {\em
  local} isometries, i.e. they only determine its universal cover ---
this cover is the simply-connected manifold $\cM_0$ used in the
standard formulation. This observation implies that one can consider
generalized supergravity models in which $\cM$ is a smooth quotient of
$\cM_0$ through the action of a discrete subgroup $\Gamma$ of the
effective\footnote{In $\cN=2$ theories, $G_\eff$ is a discrete
  quotient of $G_0$ which acts effectively on the scalar manifold,
  while for $\cN\geq 3$ theories we have $G_\eff=G_0$.} U-duality
group $G_\eff$, in which case $\pi_1(\cM)\simeq \Gamma\subset
G_\eff$. It is such generalized models that admit geometric U-fold
solutions. We will see that such U-folds are glued using U-duality
transformations belonging to $\Gamma$, so they can be lifted to string
theory U-folds only when $\pi_1(\cM)$ is a subgroup of the discrete
U-duality group $G_0(\mathbb{Z})\subset G_0$ which survives
\cite{Hull:1994ys} in string theory.

We point out a close relation between geometric U-folds and certain
moduli spaces of flat connections defined on the scalar manifold. The
latter lead to character varieties that, to our best knowledge, have
not been systematically studied in the literature. Finally, we
illustrate our construction with two examples. The first is the
``generalized axion-dilaton model'', namely $\cN=2$ supergravity
coupled to a single vector multiplet with $\cM$ given by a (generally
non-compact) Riemann surface of genus $g\geq 2$ endowed with its
uniformizing metric. The second is $\cN=8$ supergravity with scalar
manifold given by a double coset $\Gamma\backslash
\E_{7(7)}/(\SU(8)/\mathbb{Z}_{2})$, where $\Gamma$ is a discrete
subgroup of $\E_{7(7)}$. For the generalized axion-dilaton model, we
construct explicit geometric U-folds which are similar to those of
\cite{CosmicStrings}, being sourced by cosmic strings.

The construction outlined in this note leads to various questions for
further research. First of all, we lack the global mathematical formulation of four-dimensional supergravity on topologically non-trivial four-manifolds. Obtaining such formulation, aside from a mathematically interesting problem, is a mandatory step in order to understand the global structure of supergravity solutions, and in particular supergravity U-folds. Work in this direction is already in progress \cite{Lazaroiu:2016iav} and in fact the project has evolved into a long term research program focused on understanding the global mathematical structure of supergravity. Aside from studying the global structure of particular supergravity U-fold solutions, it would be interesting to properly define and study the moduli space of geometric U-folds in order to understand the space of inequivalent U-folds that can be obtained as solutions of a particular supergravity theory. This problem would require studying the relevant character varieties of discrete subgroups of U-duality groups. It would also be interesting to
consider similar constructions in $\cN=1$ supergravity and to
systematically analyze source-full U-fold solutions.  Finally, it
would be interesting to construct further explicit examples of
geometric U-folds and to study their properties and physics
consequences.

\section{The global formulation of extended four-dimensional supergravity theories}
\label{sec:Nb2Sugra}

In this section we outline the global formulation of un-gauged
$\cN\geq 2$ supergravity\footnote{For $\cN=2$, we consider only the
  theory coupled to vector multiplets.} following references
\cite{Andrianopoli:1996cm,Andrianopoli:1996ve}, describing all non-scalar fields
of the theory as global sections of appropriate fiber bundles.

Consider a four-dimensional, oriented, Lorentzian spin manifold
$(M,g)$, whose bundle of complex chiral spinors we denote by $S$.  
The {\em standard} four-dimensional supergravity theories are
constructed using certain {\em simply-connected} Riemannian scalar
manifolds $(\cM_0,\cG_0)$ which are summarized in Table \ref{table:cosets1}. 
\begin{table}[htpb!]
\centering
\begin{tabular}{|c|c|c|c|c|c|}
\hline $\cN$ & $\cM_0$ & $\dim_\R \cM_0$ & indecomposable\\ 
\hline\hline 
$2$ & PSK & $2n_v$ & not necessarily \\
\hline 
$3$ & $\frac{\SU(3,n_v)}{\mS[\U(3) \times \U(n_v)]}$  & $6n_v$ & yes \\ 
\hline 
$4$ & $\frac{\SL(2,\mathbb{R})}{\U(1)} \times \frac{\SO_0(6, n_v)}{\SO(6) \times \SO(n_v)}$  & $2+6n_v$ & no \\ 
\hline 
$5$ & $\frac{\SU(1,5)}{\mS[\U(1) \times \U(5)]}$ & $10$ & yes \\ 
\hline 
$6$ & $\frac{\SO^\ast(12)}{\U(6)}$ & $30$ & yes \\
\hline 
$8$ & $\frac{\E_{7(7)}}{\SU(8)/\Z_2}$  & $70$ & yes \\ 
\hline
\end{tabular}
\vskip 0.2in
\caption{The simply-connected scalar manifolds $\cM_0$ of standard
  four-dimensional supergravity theories, where $n_v$ denotes the
  number of vector multiplets. For $\cN\geq 3$, these scalar manifolds
  are diffeomorphic with $\R^{\dim \cM_0}$ and hence contractible.  In
  the table, $\SO_0(6,n_v)$ denotes the connected component of the
  identity in the group $\SO(6,n_v)$ (which has two connected
  components). The abbreviation PSK means a (simply-connected)
  projective special K\"ahler manifold.}
\label{table:cosets1}
\end{table}
The U-duality group $G_0$ of these theories is summarized in Table
\ref{table:cosets2}.  The Lagrangian contains a total number $n$ of
$\U(1)$ gauge fields, whose transformation under U-dualities is
determined by a certain group morphism $\rho:G_0\rightarrow
\Sp(2n,\R)$. We have $n=n_{v}+m$, where $n_v$ is the number of vector
multiplets (which can be non-zero only for $\cN\in \{2,3,4\}$ while
$m$ is the number of $\U(1)$ gauge fields in the gravity
multiplet. Namely:
\begin{enumerate}
\itemsep 0.0em
\item When $\cN=2$, $\cM_0$ is a simply-connected projective special
  K\"ahler (PSK) manifold \cite{deWit:1983rz,Freed:1997dp, ACD} and $G_0$ is a
  discrete cover of its group of special isometries\footnote{Those
    isometries of $\cM_0$ which preserve the complex structure as well
    as the flat symplectic connection (sometimes called ``duality
    symmetries'' or ``duality invariances'' \cite{deWit:1992wf}).}
  $\Iso_s(\cM_0)$. The dimension of $\cM_0$ (as a real manifold)
  equals $2n_v$ and we have $n=n_v+1$, the supplementary $\U(1)$ gauge
  field being the graviphoton.
\item When $\cN\geq 3$, $\cM_0$ is a certain simply-connected globally
  symmetric space of non-compact type, which is de Rham irreducible
  except for $\cN=4$ and $n_v\geq 1$, in which case it has
  two simply-connected and irreducible factors of non-compact type. By
  the Hadamard-Cartan theorem, it follows that $\cM_0$ is
  diffeomorphic with $\mathbb{R}^{\dim \cM_0}$.  Moreover, the
  U-duality group $G_0$ is the connected component
  $\Iso_0(\cM_0,\cG_0)$ of the identity in the isometry group
  $\Iso(\cM_0,\cG_0)$ (see Table \ref{table:cosets2}). The U-duality
  group acts transitively on $\cM_0$ and we have $\cM_0=G_0/H_0$,
  where $H_0\subset G_0$ is the isotropy group of this action.  For
  $\cN\in \{3,4\}$, the pure supergravity theory is coupled to $n_v$
  vector multiplets and $\cM_0$ is uniquely determined by $\cN$ and
  $n_v$. For $\cN\geq 5$, the theory does not admit coupling to vector
  multiplets, thus $n_v=0$, $n=m$ and $\cM_0$ is uniquely determined
  by $\cN$.
\end{enumerate}
\begin{table}[htpb!]
\centering
\begin{tabular}{|c|c|c|c|c|}
\hline $\cN$ & $m$ & $\cM_0$ & $H_0$ & $G_0$ \\ 
\hline\hline 
$2$  & $1$ & PSK & $\U(1)$ & cover of $\Iso_s(\cM_0)$\\
\hline 
$3$  & $3 $ & $G_0/H_0$  & $\mS[\U(3)\!\times\! \U(n_v)]$ & $\SU(3,n_v)$  \\ 
\hline 
$4$ & $6 $ & $G_0/H_0$  & $\U(1)\times \SO(6)\!\times\! \SO(n_v)$ & $\SL(2,\R)\!\times\! \SO_0(6,\!n_v)$ \\ 
\hline 
$5$ & $10 $ & $G_0/H_0$ & $\mS[\U(1) \!\times\! \U(5)]$ &$\SU(1,\!5)$ \\ 
\hline 
$6$ & $16 $ & $G_0/H_0$ & $\U(6)$ & $\SO^\ast(12)$\\
\hline 
$8$ & $56$ & $G_0/H_0$  & $\SU(8)/\Z_2$ & $\E_{7(7)}$ \\ 
\hline
\end{tabular}
\vskip 0.2in
\caption{The groups $G_0$ and $H_0$. The symbol $n_v$ denotes the
  number of vector multiplets while $m$ denotes the total number of
  $\U(1)$ gauge fields in the gravity multiplet. We have $n=n_v+m$.}
\label{table:cosets2}
\end{table}
\noindent For $\cN=2$ theories, the U-duality group $G_0$ is a cover
of $\Iso_s(\cM_0)$ and hence the action of $G_0$ on $\cM$ induced
through the covering map $G_0\rightarrow \Iso_s(\cM)$ may be
non-effective (we will see an example of this in Section V). Define
the {\em effective U-duality group} to be the group
$G_\eff=\Iso_s(\cM)$ of special isometries, which acts effectively on
$\cM_0$. For $\cN\geq 3$ theories, we set $G_\eff=G_0$.

In this paper, we will work with the more general choice of scalar
manifold:
\ben
\label{cM}
\cM = \Gamma\backslash\cM_0~~,
\een
where $\Gamma\subset G_\eff$ is a discrete subgroup of the effective
U-duality group such that $\Gamma\backslash\cM_0$ is smooth. We endow
$\cM$ with the metric $\cG$ induced from $\cG_0$ and let
$\pi:\cM_0\rightarrow \cM$ denote the canonical projection. Then
$(\cM_0,\cG_0)$ is the Riemannian universal cover of $\cM$ and
$\Gamma=\Aut(\pi)\simeq \pi_1(\cM)$ is the deck group of this
cover. Moreover:
\begin{enumerate}
\itemsep 0.0em
\item When $\cN=2$ and $\Gamma$ is non-trivial, the manifold $\cM$ is
  projective special K\"ahler.
\item When $\cN\geq 3$ and $\Gamma$ is non-trivial, the manifold $\cM$
  is a {\em locally} symmetric space.
\end{enumerate}
Let $\Iso(\cM,\cG)$ denote the isometry group of $(\cM,\cG)$ and
$K_\Gamma$ be the largest subgroup of $\Iso(\cM_0,\cG_0)$ which
contains $\Gamma$ as its normal subgroup. Then there exists a short
exact sequence \cite{MyersSteenrod}:
\be
1\longrightarrow \Gamma\longrightarrow K_\Gamma\longrightarrow \Iso(\cM,\cG)\longrightarrow 1~~,
\ee
which can be used to determine the isometry group of $(\cM,\cG)$.
The bosonic Lagrangian of the theory based on the
scalar manifold \eqref{cM} is globally 
determined (up to a discrete ambiguity described below) by:
\begin{itemize}
\itemsep 0.0em
\item A principal $H_0$-bundle $P$ over $\cM$.
\item A flat principal $G_0$-bundle $Q$ over $\cM$.
\end{itemize}
Namely:
\begin{enumerate}
\itemsep 0.0em
\item When $\cN=2$, we have $H_0=\U(1)$ and $P$ is the canonical
  circle bundle of $\cM$ (the circle bundle of that holomorphic
  line bundle whose first Chern class equals the K\"ahler class).
\item When $\cN\geq 3$, $H_0$ is the isotropy group of the symmetric
  space $\cM_0$, while $P$ be the principal $H_0$-bundle
  $\Gamma\backslash G_0 \rightarrow \cM$.
\end{enumerate}
When $\Gamma=1$, the corresponding bundles (which are topologically trivial) 
are denoted by $P_0$ and
$Q_0$ and are the bundles used in the standard theory. Notice that
$Q_0$ is trivial as a flat principal bundle since $\cM_0$ is
simply-connected, so $Q_0$ is determined by $\cM_0$ and $G_0$ up to
isomorphism of flat principal bundles.  In the general theory (when
$\Gamma$ is non-trivial), the flat connection of $Q$ defines the
holonomy representation:
\be
\Delta\colon \Gamma\simeq \pi_1(\cM,y) \to G_0~~, 
\ee
where  $y\in\cM$ is an arbitrary point. 
The universal cover $\cM_0$ can be viewed as a principal bundle $C$
over $\cM$ with discrete structure group given by $\Gamma$. Then $Q$
is isomorphic with the principal $G_0$-bundle $C\times_\Delta G_0$
associated to $C$ through $\Delta$.  Consider the flat vector bundle
$\cS=Q\times_\rho\bR^{2n}$ of rank $2n$ over $\cM$, which is
associated to $Q$ through the representation $\rho$. Then $\cS$ is
isomorphic with the vector bundle $C\times_{\rho\circ \Delta} \R^{2n}$ 
associated to $C$ through the representation:
\be
\rho\circ \Delta\colon \Gamma\to \Sp(2n,\bR)~~.
\ee
The bosonic fields appearing in the Lagrangian are:
\begin{itemize}
\itemsep 0.0em
\item The Lorentzian metric $g$ of $M$.
\item A smooth map $\Phi\colon M\to \cM$.  Using this map we can pull
  back $P$, $Q$ and $\cS$ to the following bundles defined over $M$:
\be
P_{\Phi} \eqdef \Phi^{\ast} (P)~~,~~ Q_{\Phi} \eqdef \Phi^{\ast} (Q)~~,~~\cS_{\Phi}\eqdef
\Phi^{\ast} (\cS)~~.
\ee
\item An $\cS_\Phi$-valued closed 2-form $F\in
  \Omega^{2}_{\cl}(M, \mathcal{S}_{\Phi})$, which describes the electric
  and magnetic field strengths of the $\U(1)$ gauge fields.
\end{itemize}
Notice that $Q_\Phi$ is a flat principal $G_0$-bundle defined over
$M$, whose holonomy representation is given by:
\be
\Delta\circ \Phi_\ast:\pi_1(M,x)\rightarrow G_0~~,
\ee
where $\Phi_\ast:\pi_1(M,x)\rightarrow \pi_1(\cM,y)$ is the homotopy
push-forward through $\Phi$ and we took $y=\Phi(x)$ for some $x\in
M$. Similarly, $\cS_\Phi\simeq Q_\Phi\rtimes_\rho \R^{2n}$ is a flat
symplectic vector bundle defined over $M$, whose holonomy
representation is given by $\rho\circ\Delta\circ \Phi_\ast$.
The fermionic field content is determined by two vector bundles:
\be
E_G = P_{\Phi}\times_{\theta_G} V_G\, , \quad E_{f} = P_{\Phi}\times_{\theta_G} V_{f}
\ee
associated to $P_{\Phi}$ through complex representations of $H_0$,
namely the gravitino representation $\theta_G\colon H_0\to \Aut(V_G)$
and the fermionic matter representation $\theta_{f}\colon H_0\to
\Aut(V_{f})$, whose precise choice depends on the theory\footnote{For certain supergravities, the spinor fields have "K\"{a}hler weight"  $1/2$, so the representations $\theta_G$ and $\theta_f$ involve  taking the square root of a $\U(1)$ sub-bundle $R$ of $P_\Phi$ which  corresponds to R-symmetry. In these cases, fermions should strictly speaking be understood as sections of vector bundles associated to the principal bundle obtained from $P_\Phi$ upon replacing $R$ with a square root $R^{1/2}$, which exists only when the first Chern class $c_1(R)$ is even and whose choice introduces a discrete ambiguity in
  the global construction of the theory; see Appendix A.} Here $V_G$
and $V_{f}$ are complex vector spaces of appropriate dimensions. The
fermionic fields in the Lagrangian are:
\begin{itemize}
\itemsep 0.0em
\item The gravitino field, which is a one-form $\Psi\in
  \Omega^{1}(M, S\otimes E_G)$ valued in the vector bundle $S\otimes
  E_G$, where $S$ is the complex spinor bundle of $M$.
\item A spinor $\chi\in \Omega^{0}(M, S\otimes E_{f})$, which is a smooth
  section of $E_{f}$.
\end{itemize}
The reader may check that the objects introduced above reproduce the
standard local (index) formulation of the field content of the theory,
including the appropriate local description of the symmetries.

\section{Geometric U-folds}
\label{sec:Ufolds}

Let $\Sol$ be a finite ordered set of fields satisfying the equations
of motion of the supergravity theory defined on $M$. Even though
$\Sol$ is defined geometrically\footnote{It is manifestly described
  using manifolds and maps of such, including sections of fiber
  bundles.}, it can in certain cases be interpreted as a U-fold when
understood by gluing local solutions defined on the sets of an open
cover of $M$. Let $\cU \eqdef \left\{ U_{a}\right\}_{a\in I}$ be an
open cover of $M$ which is a trivializing cover for both $Q_\Phi$ and
$P_\Phi$. Restricting $\Sol$ to $U_\alpha$ gives a family $\left\{
\Sol_{a}\right\}_{a\in I}$, where:
\be
\Sol_{a} \eqdef \Sol|_{U_{a}}\, , \qquad a\in I
\ee
is a solution of the theory defined on $U_a$. We are interested in how
this family glues to yield the global solution $\Sol$. For
intersecting open sets $U_a$ and $U_b$ of the cover, we have two
possibilities:
\begin{enumerate}
\itemsep 0.0em
\item The local solutions $\Sol_a$ and $\Sol_b$ are glued through
  transformations which do not involve a non-trivial U-duality.
\item The local solutions $\Sol_{a}$ and $\Sol_{b}$ are glued through
  transformations involving a non-trivial U-duality.
\end{enumerate}
If $Q_\Phi$ is topologically trivial, then we can arrange that the
first case occurs for all pairs of intersecting open sets; in this
case, we may in fact find a trivializing cover consisting of the
single open set $U=M$. In this situation, we say that $\Sol$ is {\em
  trivial as a U-fold}. If $Q_\Phi$ is topologically non-trivial then
the second possibility arises for at least one pair of intersecting
open sets of {\em any} trivializing open cover. In this case, we say
that $\Sol$ is a {\em non-trivial geometric U-fold}.  Indeed, all
fields of the theory, except for the metric and the scalar fields encoded
by $\Phi$, are either tensor fields defined on $M$ or global sections
of $\cS_\Phi$, $E_G$ or $E_{f}$. For example, the field-strength $F\in
\Omega^2_\cl(M,\cS_\Phi)$ is a closed two-form taking values in
$\cS_\Phi$. If $Q_\Phi$ is topologically non-trivial then the open
cover $\left\{ U_{a}\right\}_{a\in I}$ contains at least two
intersecting sets $U_a$, $U_b$ with a non-trivial transition function:
\be
g_{ab}\colon U_{a}\cap U_{b}\to G_0
\ee
for $Q_\Phi$. Denote by $F_{a}=F|_{U_a}\in
\Omega^2(U_a,\cS_\Phi|_{U_a})$ the field-strength of the
local solution in $U_{a}$ and by $F_{b}=F|_{U_b}\in
\Omega^2(U_b,\cS_\Phi|_{U_b})$ the field-strength of the
local solution in $U_{b}$. Then:
\be
F_{a} = (\rho\circ g_{ab}) F_{b}~~.
\ee
When $Q_\Phi$ is non-trivial, we are thus \emph{forced} to glue $F_a$
to $F_b$ using a non-trivial U-duality transformation for every
intersecting pair for which $g_{ab}$ is not identically $1$ on
$U_a\cap U_b$. Hence:
\begin{itemize}
\itemsep 0.0em
\item A smooth global solution $\Sol$ of extended supergravity for
  which $Q_{\Phi}$ is topologically non-trivial and $F$ is not
  identically zero is a non-trivial U-fold of the theory based on
  $(\cM,\cG)$.
\end{itemize}
Recall that $\Phi$ induces a map $\Phi_\ast:\pi_1(M,x)\rightarrow
\pi_1(\cM,y)$, where $x\in M$ is such that $\Phi(x)=y$.  Even when
$\cS_\Phi$ is trivial or the gauge fields vanish, solutions for which
the group $\Phi_\ast(\pi_1(M,x))$ is non-trivial can be viewed as
U-folds of the standard theory based on the scalar manifold
$(\cM_0,\cG_0)$. To see this, let $M_0$ denote the universal cover of
$M$ and $p:M_0\rightarrow M$ be the canonical projection. For any
choice of $x\in M$ and of points $x_0\in M_0$ and $y_0\in \cM_0$ such
that $p(x_0)=x$ and $\pi(y_0)=\Phi(x)=y$, the map $\Phi$ lifts to a
uniquely-determined map $\Phi_0:M_0\rightarrow \cM_0$ such that
$\Phi_0(x_0)=y_0$ and such that the following diagram commutes:
\be
\scalebox{1.0}{
\xymatrix{
M_0 \ar[d]_{p} \ar[r]^{\Phi_0} & \cM_0 \ar[d]^{\pi} \\
M   \ar[r]^{\Phi} & \cM \\
}}
\ee
Similarly, all remaining constituent fields of the solution $\Sol$
lift to fields defined on $M_0$, which together with $\Phi_0$ form a
solution $\Sol_0$ (defined on $M_0$) of the {\em standard}
supergravity theory (which is constructed using $(\cM_0,\cG_0)$).  The
original solution $\Sol$ of the theory based on $(\cM,\cG)$ can be
identified with $\Sol_0$, viewed as a {\em multivalued} solution of
the standard theory defined on $M$, with monodromies
controlled by U-duality transformations belonging to $\Gamma\subset
G_\eff$.  Indeed, let us view the universal cover
$\pi:\cM_0\rightarrow \cM$ as a principal $\Gamma$-bundle $C$ defined
over $\cM$. This pulls back through $\Phi$ to a principal
$\Gamma$-bundle $C_\Phi\eqdef \Phi^\ast(C)$ defined over $M$, which in
turn pulls back through $p$ to a principal $\Gamma$-bundle
$C'_\Phi\eqdef p^\ast(C_\Phi)$ defined over $M_0$. The map $\Phi_0$
can now be viewed as a section ${\hat \Phi}$ of $C'_\Phi$, i.e. as a
multivalued global section\footnote{By definition, a multivalued
  global section of a fiber bundle $F\rightarrow M$ is an ordinary
  global section of the bundle $p^\ast(F)\rightarrow M_0$.} of
$C_\Phi$.  This multivalued section is one-valued (i.e., it descends
to an ordinary global section of $C_\Phi$) only when $\Phi_0$ factors
through $p$, i.e. when $\Phi$ lifts to a map from $M$ to $\cM_0$. In
turn, this happens iff $\Phi_\ast(\pi_1(M,x))=1$.  When
$\Phi_\ast(\pi_1(M,x))\neq 1$, the multivalued section ${\hat \Phi}$
has monodromy valued in the structure group $\Gamma$ of $C_\Phi$,
which is a subgroup of the effective U-duality group $G_\eff$. Thus:
\begin{itemize}
\itemsep 0.0em
\item A smooth global solution $\Sol$ of the supergravity theory based
  on the scalar manifold $(\cM,\cG)$ which has the property that
  $\Phi_\ast(\pi_1(M,x))\neq 1$ is a non-trivial geometric U-fold.
\end{itemize}
The condition $\Phi_\ast(\pi_1(M,x))\neq 1$ requires that both $M$ and
$\cM$ have non-trivial first homotopy group and that $\Phi$ be a
homotopically non-trivial map. Similarly, the flat principal bundle
$Q_\Phi$ (and hence also the flat vector bundle $\cS_\Phi)$ is trivial
unless the holonomy representation $\Delta\circ
\Phi_\ast:\pi_1(M)\rightarrow G_0$ is non-trivial, which requires
$\Phi_\ast(\pi_1(\cM))\neq 1$ and that the group morphism $\Delta$ be
non-trivial (i.e. that $Q$ be a non-trivial flat principal bundle over
$\cM$). In particular:
\begin{itemize}
\itemsep 0.0em
\item Every smooth global solution $\Sol$ of a standard extended
  supergravity theory (with simply-connected scalar manifold $\cM_0$)
  is trivial as a U-fold, as is any solution of the generalized theory
  based on $\cM$ whose underlying space-time is simply-connected or
  whose scalar field configuration $\Phi$ is homotopically trivial.
\end{itemize}

\section{Pre-classifying geometric U-folds}
\label{sec:moduliUfolds}

\noindent Let:
\beqa
&& \fM_{G_0}(\cM)= \mathrm{Hom}\left(\pi_1(\cM),G_0\right)/G_0\nonumber\\
&& \fM_{G_0}(M)= \mathrm{Hom}\left(\pi_1(M),G_0\right)/G_0
\eeqa
denote the moduli spaces of flat principal $G_0$-bundles over $\cM$ and
$M$ respectively, where the quotient is through the adjoint action of
$G_0$. A smooth map $\Phi:M\rightarrow \cM$ is called {\em admissible}
if there exist fields defined on $M$ which, together with $\Phi$, form
a solution $\Sol=(\Phi,\ldots)$ of the theory defined on $M$. Let
$\cC^{\infty}_{\ad}(M,\cM)$ denote the space of admissible maps from
$M$ to $\cM$. The homotopy space $[M,\cM]_\ad$ of admissible maps is
the space of connected components of $\cC^{\infty}_{\ad}(M,\cM)$ with
respect to the natural topology and can be obtained upon dividing through the
homotopy equivalence relation $\sim$:
\be
[M,\cM]_\ad \eqdef \pi_0(\cC^{\infty}_{\ad}(M,\cM))=\cC^{\infty}_{\ad}(M,\cM)/\sim~~.
\ee
Since the isomorphism class of $Q_\Phi\eqdef \Phi^\ast(Q)$ as a
flat bundle depends only on the homotopy class of $\Phi$ and on the
isomorphism class of $Q$ as a flat bundle, the pull-back operation
induces a well-defined map:
\ben
\label{pb}
[M,\cM]_\ad\times \fM_{G_0}(\cM) \ni ([\Phi],[Q])\rightarrow [\Phi^\ast(Q)]\in \fM_{G_0}(M)~~,
\een
whose image we denote by $\fM_{G_0}^\ad(M)$ and call the {\em pre-moduli
  space} of geometric U-folds. The moduli space $\fM(M)$ of geometric
U-folds on $M$ (if properly defined) should map to $\fM_{G_0}^\ad(M)$
with fiber given by those geometric U-folds which have isomorphic
bundles $Q_\Phi$. Effectively describing $\fM(M)$ and $\fM_{G_0}^\ad(M)$
appears to be a formidable problem, given the complicated nature of
the equations of motion of the theory. Notice that \eqref{pb} is
controlled by the homotopy push-forward:
\be
[M,\cM]_\ad\times \pi_1(M,x) \ni ([\Phi],[\alpha])\rightarrow \Phi_\ast([\alpha])\in \pi_1(\cM,y)~~,
\ee
where $\Phi_\ast([\alpha])=[\Phi \circ \alpha]$.

\section{$\cN=2$ supergravity coupled to one vector multiplet}
\label{sec:N2Ufolds}

The \emph{standard axion-dilaton model} is $\mathcal{N}=2$
supergravity coupled to one vector multiplet with simply-connected scalar
manifold given by the (open) upper-half plane\footnote{Since
  $\U(1)/\Z_2$ is isomorphic with $\U(1)$ through the isogeny
  $z\rightarrow z^2$, while $\SL(2,\R)\simeq \SU(1,1)$, we also have
  the coset presentations $\H\simeq
  \SL(2,\R)/\U(1)\simeq\SU(1,1)/\U(1)$, which are not minimal since
  the group appearing in the numerator is a double cover of
  $\Iso_0(\cM_0,\cG_0)=\PSL(2,\R)$ and $\SL(2,\R)$ acts non-effectively on $\H$.}:
\be
\cM_0=\H=\PSL(2,\R)/\U(1)~~,
\ee
equipped with the rescaled Poincar\'e metric $\mathcal{G}_{0}$ of
constant Gaussian curvature $-2$ (scalar curvature $-4$). The latter
has the squared line element:
\be
\dd s_0^2=\frac{1}{2(\Im \tau)^2} \dd \tau \dd {\bar \tau}=2(\cG_0)_{\tau{\bar \tau}}\dd \tau \dd {\bar \tau}~~
\ee
and is the unique $\PSL(2,\R)$-invariant metric of the given scalar
curvature.  The connected component of the isometry group is
$\Iso_0(\cM_0,\cG_0)=\PSL(2,\R)$, which is also the group of
orientation-preserving isometries. The manifold $\cM_0$ is projective
special K\"ahler\footnote{Every Riemann surface of genus $g\geq 2$ is
  projective special K\"ahler when endowed with its uniformizing
  metric \cite{Baues:1999aya}.} with the following global
prepotential defined on the conical special Kahler domain ${\cal
  D}=\{(X^0,X^1)\in \C^2|\mathrm{Re} ({\bar X^0}X^1)>0\}$:
\be
\hat{\cF}_0(X^0,X^1)= -i X^0 X^1=-(X^0)^2\cF_0(\tau)~~,
\ee
where $\tau\eqdef i\frac{X^1}{X^0}$ and $\cF_0(\tau)=\tau$. We have
$(\cG_0)_{\tau{\bar \tau}}=\frac{\partial^2 K_0}{\partial \tau \partial
  {\bar \tau}}=\frac{1}{(2\Im\tau)^2}$, where $K_0$ is the (global)
K\"ahler potential in the gauge $X^0=\frac{i}{2}$:
\be
K_0(\tau)= -\ln (\Im \tau)~~;
\ee
notice that we are using the Riemannian (positive-definite) metric on
$\cM_0$.  The effective U-duality
group is the group of special isometries
$G_\eff=\Iso_s(\cM_0)=\PSL(2,\R)$, while the U-duality group is its
double cover $G_0=\SL(2,\R)$.  The canonical circle bundle is the
trivial $\U(1)$-bundle $P_0=\cM_0\times \U(1)$. We have $n_v=1$ and
$n=2$, while the symplectic representation $\rho:G_0\rightarrow
\Sp(4,\R)$ is given (up to equivalence of representations) by:
\be
\SL(2,\R)\ni A=\left[\begin{array}{cc} a & b \\ c & d\end{array}\right]\!\rightarrow\! \rho(A)=\left[\begin{array}{cc} a I_2 & b\Theta_2 \\ c\Theta_2 & d I_2\end{array}\right]~~,
\ee
where $I_2=\left[\begin{array}{cc} 1 & 0 \\ 0 & 1\end{array}\right]$
and $\Theta_2\eqdef \left[\begin{array}{cc} 1 & 0 \\ 0 &
    -1\end{array}\right]$.  This determines the trivial rank four flat
symplectic vector bundle $\cS_0=\cM_0\times \R^4$.

Let $\Gamma\subset G_\eff=\PSL(2,\R)$ be a Fuchsian group without elliptic
elements. The {\em generalized axion-dilaton model} determined by
$\Gamma$ is $\cN=2$ supergravity coupled to a single vector multiplet
with smooth scalar manifold:
\ben
\label{unif}
\cM = \Gamma\backslash\H~~,
\een
endowed with the constant negative curvature metric $\cG$ induced by
$\cG_0$. Thus $\cM$ is a (possibly non-compact) smooth Riemann surface of
genus $g\geq 2$ while $\cG$ is its (rescaled) uniformizing metric (the
unique metric on $\cM$ which has constant Gaussian curvature equal to
$-2$). By the uniformization theorem, any smooth Riemann surface of
genus $g\geq 2$ endowed with its uniformizing metric can be presented
as in \eqref{unif}.  Notice that $\cM$ has finite volume when $\Gamma$
is co-finite and that it is compact when $\Gamma$ is co-compact. When
endowed with the complex structure $J$ induced from $\H$, the
Hermitian manifold $(\cM,J,\cG)$ is projective special K\"ahler. Any group
morphism $\Delta:\Gamma\rightarrow G_0=\SL(2,\R)$ determines a flat 
principal $\SL(2,\R)$-bundle $Q=C\rtimes_\Delta \SL(2,\R)$ and a
rank four flat symplectic vector bundle $\cS=C\rtimes_{\rho\circ \Delta} \R^4$
with monodromy representation $\rho\circ \Delta$.  The moduli space of
flat $\SL(2,\R)$-bundles on $\cM$ is the well-studied character variety:
\be
\!\!\!\!\!\!\fM_{\SL(2,\R)}(\cM) = \Hom(\Gamma,\SL(2,\R))/\SL(2,\R)~~,
\ee
which is closely related to the Teichm\"{u}ller space of $\cM$. 
This model admits non-trivial geometric U-folds, which can be promoted
to string theory backgrounds only when $\Gamma$ is a subgroup of
$\PSL(2,\Z)$. Below, we construct examples of such U-folds. 

\subsection{Example: fluxless axion-dilaton U-folds}

Consider the generalized axion-dilaton model defined by a Fuchsian
group $\Gamma\subset \PSL(2,\R)$ without elliptic elements. Focusing
on solutions for which the two gauge field strengths vanish
identically, we can truncate the bosonic part of the action to:
\begin{equation}
S[g,\tau] = \int_{M}\left\{ \ast R + \frac{\dd\tau\wedge \ast \dd\tau}{2(\Im \tau)^2}\right\}~~.
\end{equation}
Take the space-time manifold to be of the form: 
\be
M=\R^2\times \Sigma~~,
\ee
where $\Sigma$ is an oriented connected surface without boundary which admits a
(possibly-incomplete) flat Riemannian metric $g_2$ and take $g$ to be a
flat Lorentzian metric of the form $g=\eta_2\times g_2$, where
$\eta_2$ is the Minkowski metric on $\R^2$. Further, assume that
$\tau$ does not depend on the coordinates of $\R^2$, so that it can be
viewed as a smooth map $\tau:\Sigma \rightarrow \cM$. Then the
Einstein equation is satisfied while the equation of motion for $\tau$
reduces to:
\begin{equation}
\label{eqtau}
\partial\bar{\partial}\tau + \frac{\partial\tau\bar{\partial}\tau}{\tau\bar{\tau}} = 0~~,
\end{equation}
where $\partial$ and ${\bar \partial}$ are the Dolbeault differentials
defined by the complex structure of $\Sigma$ corresponding to the
conformal class of $g_2$. A particular class of solutions of
\eqref{eqtau} is given by maps $\tau$ which satisfy ${\bar
  \partial}\tau=0$ and hence are holomorphic on $\Sigma$. As explained
in Section III, these can be viewed as multivalued holomorphic maps
${\hat \tau}$ from $\Sigma$ to $\H$ whose monodromy representation takes
values in $\Gamma$ and hence involves U-duality transformations. Let
$\tau_0:\Sigma_0\rightarrow \H$ denote the lift of $\tau$ at a point
$x_0\in M_0$, where $\Sigma_0$ is the universal cover of ${\Sigma}$.
The universal cover of $M$ is $M_0=\R^2\times \Sigma_0$. Distinguish
the cases:
\begin{enumerate}
\itemsep 0.0em
\item $(\Sigma,g_2)$ is complete. Then $(\Sigma,g_2)$ is an oriented
  Euclidean space-form and hence must be the Euclidean plane
  (conformally, the complex plane $\C$), the flat infinite cylinder
  (conformally, the complex punctured plane $\C\setminus \{0\}$) or a
  flat torus (conformally, an elliptic curve). Since $\cM$ has genus
  at least two, the Picard theorem for Riemann surfaces forces $\tau$
  to be constant, so such solutions are trivial as
  U-folds. 
\item $(\Sigma,g_2)$ is incomplete. Then $\Sigma$ can be any open
  domain of the Euclidean plane. This leads to non-trivial geometric
  U-folds provided that $\Sigma$ has non-trivial fundamental group.  A
  physically interesting example is $\Sigma=\C\setminus A$, where
  $A=\{p_1,\ldots,p_k\}$ is a non-empty finite set of points of the
  complex plane. In this case, the Riemannian universal cover
  $\Sigma_0$ is conformally equivalent with the complex plane or with
  the Poincar\'e disk (depending on whether $k=0,1$ or $k\geq 2$) and
  ${\hat \tau}$ can have non-trivial $\Gamma$-valued monodromies around each
  of the points $p_j$, giving a four-dimensional solution 
  similar to the cosmic string of \cite{CosmicStrings}. The
  Hodge dual $H_0\eqdef \ast_M \dd \tau_0\in \Omega^3(M_0)$ satisfies
  $\dd H_0=0$ since $\tau_0$ is holomorphic. Thus $H_0=\dd B_0$ for a
  globally-defined two-form $B_0\in \Omega^2(M_0)$. The standard
  supergravity theory with scalar manifold $(\cM_0,\cG_0)$ admits
  cosmic strings which couple to this potential. Then the solution
  $\tau$ can be interpreted as being sourced by strings with
  worldvolume $\R^2\times \{p_j\}$, which are responsible for the
  monodromies of ${\hat \tau}$.
\end{enumerate}
The supergravity backgrounds constructed above can be lifted to string
theory only when $\Gamma$ is a subgroup of $\PSL(2,\Z)$.

Unlike our solutions, the backgrounds of \cite{CosmicStrings} arise in
ten-dimensional IIB supergravity. Also notice that the construction of
loc. cit. uses the Fuchsian group $\PSL(2,\Z)$, which contains
elliptic elements. As a consequence, the quotient
$\PSL(2,\Z)\backslash \H$ (endowed with the constant negative
curvature metric induced by $\cG_0$) is a projective special K\"ahler
{\em orbifold} (topologically, a once-punctured sphere with two
conical orbifold points of orders $2$ and $3$) which coincides with
the moduli space of elliptic curves endowed with its Weil-Petersson
metric. This orbifold is in principle not an admissible scalar manifold in
supergravity, so the construction of \cite{CosmicStrings} makes
physics sense only in non-perturbative IIB string theory, where it
gives an F-theory background. By contrast, the construction above
works classically in four-dimensional $\cN=2$ supergravity
(where it produces solutions containing cosmic string sources), since it
involves smooth target manifolds for the scalar field $\tau$.

\section{$\cN= 8$ supergravity}
\label{sec:N8Ufolds}

The scalar manifold of \emph{standard} $\cN=8$ supergravity is given by:
\be
\cM_0 = \E_{7(7)}/(\SU(8)/\Z_{2})~~.
\ee
In this
case, $H_0=\SU(8)/\mathbb{Z}_{2}$ is the maximal compact sub-group of
$\E_{7(7)}$ and $P_0$ is the principal $H_0$-bundle given by the
canonical projection $\E_{7(7)}\rightarrow \cM_0$. The U-duality group
is $G_0=\E_{7(7)}$ and $\rho$ is the $56$-dimensional representation
of $G_0$. The U-duality group has the following polar decomposition
\cite{2009arXiv0902.0431Y}:
\be
\E_{7(7)}\simeq H_0\times \R^{70}~~.
\ee
Since $\pi_1(H_0)=\Z_2$, this gives $\pi_1(\E_{7(7)}) =
\Z_2$ and shows that $\cM_0$ is diffeomorphic with
$\R^{70}$, hence contractible.  Thus every fiber bundle on $\cM_0$ is
topologically trivial and the standard theory does not admit
non-trivial geometric U-folds.

Let $\Gamma$ be a discrete subgroup of $\E_{7(7)}$. Then any smooth 
Clifford-Klein form:
\be
\label{eq:generalN8}
\cM = \Gamma \backslash\cM_{0}=\Gamma \backslash \E_{7(7)}/(\SU(8)/\Z_2)~~
\ee 
is admissible as a scalar manifold of $\mathcal{N}=8$ supergravity.
At least when $\cM$ is non-compact, we expect such generalized models
to admit geometric U-fold solutions, which could be promoted to string
theory backgrounds when $\Gamma$ is a subgroup of
$\E_{7(7)}(\mathbb{Z})$. The pre-moduli space of such geometric
U-folds is controlled by the character variety:
\be
\fM_{\E_{7(7)}}(\cM)= \Hom(\Gamma,\E_{7(7)})/\E_{7(7)}\, ,
\ee
which, to our best knowledge, was not systematically studied in the 
supergravity literature.

\section{Final remarks}

In this note, we have considered supergravity U-folds from the point of view of the generic ungauged four-dimensional (Lorentzian) effective theory that could potentially correspond to the effective theory describing string theory compactified on certain locally geometric U-fold compactification backgrounds. This is in contrast to the scenario usually explored and studied in the literature, which focuses on approaching U/T-fold compactification backgrounds from the ten-dimensional string/supergravity point of view as well as the local structure of the corresponding effective gauged supergravity, see for example \cite{DallAgata:2007egc,Condeescu:2013yma,Andriot:2012an} and references therein. Therefore, it would be useful to understand in detail the relation between the set up we consider in this note and the higher-dimensional constructions already present in the literature, see for example
\cite{Kumar:1996zx,Hellerman:2002ax,Marchesano:2007vw,Wecht:2007wu,ReidEdwards:2008rd,deBoer:2010ud,McOrist:2010jw}. In this note we do not explicitly address the way in which U-fold compactification backgrounds are constructed from the ten-dimensional point of view. Instead, we focus on the global structure of the generic bosonic sector of four-dimensional supergravity, which should correspond to the effective theory describing some (certainly not all) string compactifications on certain locally geometric U-fold backgrounds. In practical terms, the way we do this is by imposing global consistency with the global U-duality transformations which act on the fields of the local theory. In other words, we glue local standard supergravities in four dimensions by using U-duality transformations. These are global symmetries of the local equations of motion and hence can be used to define a consistent globally defined theory whose solutions are local solutions of standard supergravity glued by U dualities. 

It is not immediately clear how to relate in detail the construction proposed in this note to the explicit way in which some U/T-fold compactification backgrounds are constructed. What we can say is that ten-dimensional U dualities (such as the ones used to construct T/U-fold compactification backgrounds) correspond to isometries of the scalar manifold of the four-dimensional theory, which act in a symplectic representation on the gauge fields and their duals. The isometry group acts in a very non-linear way on the scalar manifold of the theory, which corresponds to the moduli space of the compactification, and hence contains information about the \emph{extra dimensions} and the way the theory has been obtained from ten dimensions through compactification on a U-fold compactification background. In order to find a direct relationship between the way a U-fold compactification background is constructed and the twisting (as described in our manuscript) of the corresponding four-dimensional effective theory, one should actually perform the compactification and find the global structure of the corresponding effective theory. To the best of our knowledge, this has not been studied in the literature. This topic deserves further attention and we plan to address it in the future.

\begin{acknowledgments}  
We thank Tom\'as Ort\'in for participating during the early stages of
this project. We also thank Vicente Cort\'es, Vicente Mu\~noz and
Diego Regalado for correspondence. We also would like to thank the two anonymous referees from \emph{Journal of Physics A: Mathematical and Theoretical} for suggestions and comments that have improved the manuscript. The work of C.I.L. is supported by grant IBS-R003-S1. The work of C.S.S. is supported by the ERC Starting
Grant 259133 – Observable String.
\end{acknowledgments} 

\appendix

\section{Square root ambiguity of spinor fields}
\label{app:Z2Ufolds}

As mentioned above, for certain supergravities the representations
$\theta_G$ and $\theta_f$ used in the construction of $E_G$ and $E_f$
involve taking the square root of the fundamental representation of a
$\U(1)$-sub-bundle of $P_{\Phi}$, a procedure which generally is
obstructed and non-unique. For example, the spinors of $\mathcal{N}=2$
supergravity are properly-speaking valued in complex vector bundles
associated to a square root $P^{1/2}_{\Phi}$ of the $\U(1)$-bundle
$P_{\Phi}$. Thus $c_{1}(P_{\Phi}) = \Phi^{\ast}c_{1}(P)$ must be
even. The square roots of $P_\Phi$ have first Chern classes lying in
the preimage of $c_1(P_\Phi)$ through the endomorphism of $H^2(M,\Z)$
given by muMcOrist:2010jwltiplication with $2$. This phenomenon does not seem to
have been studied systematically in the supergravity literature.

\phantomsection
\bibliography{References}
\label{biblio}
\newpage  
\end{document}